\documentclass[11pt]{article}
\usepackage{float}
\usepackage[english]{babel}
\usepackage[utf8x]{inputenc}

\usepackage{textcomp}
\usepackage{url}	
\usepackage{tcolorbox}

\usepackage[T1]{fontenc}
\usepackage{mathpazo}
\usepackage[margin=3cm]{geometry}
\usepackage{algorithm2e}

\usepackage{amsfonts}
\usepackage{amsmath}
\usepackage{amssymb}
\usepackage{latexsym}
\usepackage{mathtools}
\usepackage{physics}

\usepackage{xcolor}
\definecolor{RED}{rgb}{1,0,0}

\usepackage[pdfencoding=auto, psdextra]{hyperref}
\usepackage{bookmark}%

\usepackage{hyperref}
\hypersetup{pdfpagemode=UseNone}
\usepackage{amsthm}
\usepackage{cleveref}
\usepackage{authblk}

\theoremstyle{definition}

\newcommand{\tbd}[1]{{{\color{blue}}}}

\begin{document}
\title{Shallow-circuit Supervised Learning on a Quantum Processor}
\author[2]{Luca Candelori}
\author[1,*]{Swarnadeep Majumder}
\author[1]{Antonio Mezzacapo}
\author[1]{Javier Robledo Moreno}
\author[2]{Kharen Musaelian}
\author[2]{Santhanam Nagarajan}
\author[2,*]{Sunil Pinnamaneni}
\author[1]{Kunal Sharma}
\author[2]{Dario Villani}

\renewcommand{\Affilfont}{\small}

\affil[1]{{\em IBM Quantum, IBM T.J. Watson Research Center, Yorktown Heights, NY 10598, United States}}
\affil[2]{{\em Qognitive, Inc., 119 W $24^{th}$ St., New York, NY \ 10011, United States}}

\affil[*]{\em Primary authors - These authors contributed equally to this work}
\maketitle

\begin{abstract}
Quantum computing has long promised transformative advances in data analysis, yet practical quantum machine learning has remained elusive due to fundamental obstacles such as a steep quantum cost for the loading of classical data and poor trainability of many quantum machine learning algorithms designed for near-term quantum hardware.
In this work, we show that one can overcome these obstacles by using a linear Hamiltonian-based machine learning method which provides a compact quantum representation of classical data via ground state problems for k-local Hamiltonians. We use the recent sample-based Krylov quantum diagonalization method to compute low-energy states of the data Hamiltonians, whose parameters are trained to express classical datasets through local gradients. 
We demonstrate the efficacy and scalability of the methods by performing experiments on benchmark datasets using up to 50 qubits of an IBM Heron quantum processor.








\end{abstract}

\section{Introduction}\label{sec:introduction}
Over the past decade, rapid progress in Artificial Intelligence (AI) and parallel advances in quantum computing have fueled interest in Quantum Machine Learning (QML). The hope is that quantum processors can offer new computational primitives for tackling machine learning problems. Early work in QML was driven by the promise of superpolynomial speedups, inspired by quantum algorithms based on linear system solvers such as HHL \cite{harrow2009quantum}. These approaches suggested that, in principle, quantum computers might process certain structured data more efficiently than classical methods. However, they relied on efficient state preparation for loading classical data and efficient readout from the final quantum state, and both requirements emerged as major bottlenecks \cite{aaronson2015read}. In several cases, once classical algorithms were given comparable access models, the promised quantum advantages disappeared, leading to a series of dequantization results \cite{tang2022dequantizing}. 

Furthermore, early QML algorithms require fault-tolerant quantum computing platform. Current pre-fault-tolerant quantum computers can only allow for algorithms that rely on shallow-depth circuits ~\cite{kandala2019error,lanes2025framework, Robledo2024SKQD, yu2025QuantumCentricAlgorithmSampleBased}. This motivated the development of Variational Quantum Algorithms (VQAs), which were proposed as promising low-depth alternatives to fault-tolerant quantum algorithms. VQAs employ ansatzes in the form of shallow quantum circuits with trainable parameters, making them suitable for current quantum processors. However, designing an ansatz, or quantum neural network, with a good inductive bias for a given problem remains difficult. The optimization landscape can suffer from barren plateaus \cite{larocca2025barren} and unfavorable local minima \cite{anschuetz2022QuantumVariationalAlgorithms}, which makes  training quantum models challenging. Recent work has introduced parameterized circuit families that incorporate measurement and feedforward operations \cite{deshpande2024dynamic}, although their implementation on quantum processors is still limited. Quantum kernel methods provide another direction for QML by leveraging quantum feature maps to perform the kernel trick \cite{havlicek2019SupervisedLearningQuantumenhanced}. In principle, this approach can yield separation between classical and quantum learners \cite{liu2021rigorous}, but their broad applicability hinges on identifying a kernel with the right inductive bias for realistic datasets. For a recent large-scale demonstration of quantum kernel methods, we refer the reader to \cite{agliardi2024mitigating}.

Practical machine learning on current quantum processors therefore requires approaches that efficiently embed classical data, avoid optimization collapse, and remain robust to hardware noise.  Another key requirement is that the circuits used should not be efficiently classically simulable.  We tackle these issues in this work using a linear Hamiltonian-based machine learning framework developed by Qognitive Inc. and inspired by Quantum Cognition Machine Learning (QCML) \cite{musaelian2024QuantumCognitionMachine, candelori2025RobustEstimationIntrinsic}, which has been successfully applied to finance \cite{samson2024QuantumCognitionMachine, rosaler2025SupervisedSimilarityHighyield}, healthcare \cite{caro2025QuantumCognitionMachine}, and pure data science \cite{candelori2025RobustEstimationIntrinsic, abanov2025QuantumGeometryData}.  Its key innovation is the learning of data encoding as a Hamiltonian constructed from feature operators and the feature values for any data point. As a result, each data point induces a distinct Hamiltonian, even though the underlying feature operators remain fixed. For a classification task, the label is then inferred by measuring an observable on the ground state associated with the Hamiltonian corresponding to each data point. This expressive embedding encodes only the relevant information, making it feasible to run on current quantum processors.  The idea of encoding data into Hamiltonians was also explored in \cite{jerbi2024shadows}, although that work employed a fixed Hamiltonian embedding rather than learning task-dependent operators.

The main learning problem in this framework is the identification of the set of feature and label operators that are aligned with the underlying learning task. To achieve that, we need to approximate ground states for Hamiltonians that corresponds to a data points, a task addressed on quantum hardware by the recently introduced sample-based Krylov quantum diagonalization (SKQD) algorithm \cite{yu2025QuantumCentricAlgorithmSampleBased}, which has provable convergence for approximating ground state energies under the assumption that the ground states are sparse. SKQD draws quantum samples from Krylov quantum states and performs a classical diagonalization in the subspace spanned by the sampled bitstrings to approximate both the ground state and the low-energy spectrum. To learn the quantum  operators associated with classical features, we begin with a fixed set of Pauli terms for each feature operator and initialize their coefficients randomly. These coefficients are then updated during training. In principle, the exact gradients depend on all eigenstates of the Hamiltonian, which is inefficient to compute at scale. 

We highlight several key properties of our approach. By embedding classical data into $k$-local Hamiltonians, we are effectively learning a compact encoding. Krylov states involved terms of the form $e^{-i H j \Delta t}$, so classical features enter as weighted combinations of Pauli terms inside the single- and two-qubit gates that generate these states. Moreover, our method operates on sparse ground states obtained from SKQD, which implies that we probe only a small subspace of the full Hilbert space. 

Our workflow is designed to scale on existing pre-fault-tolerant quantum computers. We experimentally evaluate our method on a binary classification problem with ten features and demonstrate that the model can be trained while yielding non-vanishing gradients on an IBM Heron quantum processors up to 50 qubits.

This paper is organized into four sections. Section \ref{sec:Methods} starts with subsection \ref{sec:qcml} describing the machine learning model, its representation in the Pauli string bases, and the cross-entropy loss function for classification. Subsection \ref{sec:Gradients} describes the associated gradient calculations, the SKQD subroutine, and Pauli-string transition probabilities. Subsection \ref{sec:Training} describes the hybrid quantum-classical training process. Section \ref{sec:Results} discusses the data set used, the experimental setup, and hybrid simulations on current quantum processors, and the conclusions are presented in \ref{sec:DiscussionAndNextSteps}.

\section{Methods}
\label{sec:Methods}

\begin{figure}[!htbp]
\centering
\includegraphics[]{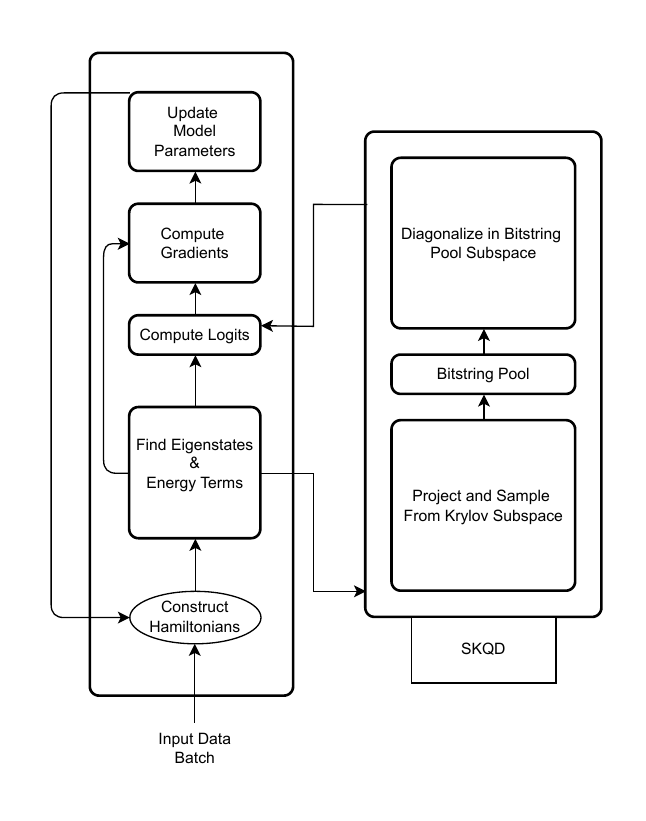}
\caption{\textbf{Model Architecture}} 
\label{fig:model_architecture}
\end{figure}

\subsection{A quantum model for classical data} \label{sec:Description}

\label{sec:qcml}
The linear Hamiltonian-based machine learning paradigm uses the mathematical framework of quantum mechanics to tackle data analysis problems. Following the notation and conventions in \cite{candelori2025RobustEstimationIntrinsic}, we introduce a set $X=t \times D$, containing $t$ data points $x_1,...,x_t$.  Each data point $x_i$ consists of a $D$-dimensional real-valued vector of data features $x_i = \left( a_i^1, \dots,a_i^D \right)$. In addition, an associated $t \times K$ set $Y$ contains $t$ targets $y_1,\dots,y_t$, each of which is a $K$-dimensional real-valued vector $y_i = \left( b_i^1, \dots,b_i^K \right)$. We consider model parameters $\mathcal{M}$ consisting of the bias operator $B$, input feature operators $\mathcal{X}=\{X_f\}_{f=1}^D$ called the {\em matrix configuration}, and output operators $\mathcal{O}=\{O_j\}_{j=1}^K$. All of these operators are Hermitian. For any state $\ket{\psi}$ in the Hilbert space, the position vector and logits then become
\begin{eqnarray}
    y(\psi) &=& \left(\expval{O_1}{\psi} , \dots, \expval{O_K}{\psi} \right) \nonumber \\
    & = & \left(l_1 , \dots, l_K\right)\in \mathbb{R}^K. \label{eq:logits}
\end{eqnarray}
For classification tasks, the model outputs $l_j$'s, called logits, are used to calculate the loss function \cite{mayraz_recognizing_2002}. For any data point $x$, the quantum state $\ket{\psi_0(x)}$ is constructed encoding not only the position of $x$ in the feature space but also its relationship with all the other points in the dataset via the Hamiltonian. The quantum state corresponding to each data point is computed as the ground state of the Hamiltonian
\begin{equation}
    H(x) = B - \sum_{f=1}^{D} a^f X_f.  \label{eq:linham}
\end{equation}
In physics and chemistry applications, one builds the model Eq.~(\ref{eq:linham}) from first principles. Here however, in the context of data analysis and machine learning, the operators $B$ and $X_f$ are learned during model training. 
The fact that the dataset is loaded through the Hamiltonian helps alleviate the data input problem typical of many QML proposals that use angle or amplitude encodings, for which one would require deep circuits. 

The matrix configuration $\mathcal{X}$, along with the bias operator $B$ and output operators $\mathcal{O}$ can be expressed in a Pauli string basis $P=\{P_p\}_{p=1}^P$ \cite{nielsen_quantum_2010} as follows
\begin{equation}
    B = \sum_{p=1}^{P} \beta^p P_p, \; X_f = \sum_{p=1}^{P} \chi_f^p P_p, \text{ and } O_k = \sum_{p=1}^{P} o_k^p P_p, \label{eq:bxo}
\end{equation}
In this work, we consider the set of 1 and 2-local Pauli strings on n qubits with the non-identity terms being adjacent. This set will be referred to as the adjacent Pauli string pattern. This is done to reduce the depth of the quantum circuits for execution on current quantum processors. In the future, one could consider different patterns and higher-weight Pauli operators.

Model training equates to learning the coefficients $\{ \beta^p, \ \chi_f^p, \ o_k^p \}$ of the operators in Eq.~(\ref{eq:bxo}) by minimizing a loss $\mathcal{L}$ as
\begin{equation}
    \{\beta^p,\ \chi_f^p, \ o_k^p \} = \text{argmin}_{\{\beta^p,\ \chi_f^p, \ o_k^p\}} \mathcal{L}, \label{eq:train}
\end{equation}
For classification tasks, this loss is computed for each data point from probabilities $p_k$ obtained by normalizing the logits in Eq. (\ref{eq:logits}) using a softmax function, along with one-hot encoded ground-truth labels $\tau_k$ as
\begin{eqnarray}
p_k &:=& \frac{\exp ( l_k )}{\sum_{k=1}^K \exp (l_k) }, \nonumber\\ 
\mathcal{L} &:=& - \sum_{k=1}^{K}\tau_k \log(p_k).
\end{eqnarray}

\subsection{Gradients}
\label{sec:Gradients}
The model training described in Eq.~(\ref{eq:train}) is achieved through back-propagation, which requires calculating the partial derivatives of the loss function $\mathcal{L}$ with-respect-to the coefficients $\{\beta^p,\ \chi_f^p, \ o_k^p\}$.  Using perturbation theory \cite{kato_perturbation_1995}, these derivatives are computed as

\begin{eqnarray}
\frac{\partial \mathcal{L}}{\partial \beta^p} &=& -2 \sum_{k=1}^{K}  \frac{\partial \mathcal{L}}{\partial l_k} \operatorname{Re} \sum_{m=1}^{2^n}\sum_{q=1}^{P}o_k^q\frac{\bra{\psi_0}P_q\ket{\psi_m}\bra{\psi_m}P_p\ket{\psi_0}}{E_m-E_0}, \nonumber \\
\frac{\partial \mathcal{L}}{\partial \chi_f^p} &=& 2 a^f \sum_{k=1}^{K}  \frac{\partial \mathcal{L}}{\partial l_k} \operatorname{Re} \sum_{m=1}^{2^n}\sum_{q=1}^{P}o_k^q\frac{\bra{\psi_0}P_q\ket{\psi_m}\bra{\psi_m}P_p\ket{\psi_0}}{E_m-E_0}, \ \text{and} \nonumber \\
\frac{\partial \mathcal{L}}{\partial o_k^p} &=& \bra{\psi_0}P_p\ket{\psi_0}\frac{\partial \mathcal{L}}{\partial l_k}
\label{eq:gradients}
\end{eqnarray}


Note that while the analytical formulas for the partial derivative with-respect-to $\beta^p$ and $\chi_f^p$ sum over exponentially many energy terms, our optimization method truncates this series after a small number of terms, which is a hyper parameter of the training workflow. 


One important part of computing the gradient in \ref{eq:gradients} is to compute the ground state and set of excited states. Here we use recently discovered Sample-based Krylov Quantum Diagonalization methods ~\cite{yu2025QuantumCentricAlgorithmSampleBased}. SKQD algorithm is a quantum-classical hybrid approach designed to approximate ground-state properties of many-body Hamiltonians using samples drawn from a Krylov basis generated by a collection of quantum circuits acting on a reference state $|\Psi_\textrm{init}\rangle$. It builds on the more general framework of ground state methods that do classical diagonalization over measurement subspaces~\cite{Robledo2024SKQD, kanno2023QSCI}, in which a quantum computer is used to generate a subspace spanned by sampled computational basis states, and a classical computer then performs the diagonalization of the Hamiltonian projected into this subspace. The key insight of SKQD is to employ \textit{time-evolution circuits}—inspired by Krylov subspace methods in classical numerical linear algebra—as the generators of sampling distributions. This allows the algorithm to systematically, and with probable convergence, explore the subspace of the Hilbert space that carries most of the weight of the true ground-state wavefunction. 

Let $H$ be an $n$-qubit Hamiltonian with ground state $\ket{\phi_0}$ and ground-state energy $E_0$. Assume the existence of a reference state $\ket{\Psi_{\text{init}}}$ that is easily preparable on a quantum computer and has non-negligible overlap with $\ket{\phi_0}$, i.e., $|{\bra{\Psi_{\text{init}} \ket{\phi_0}}}^2$ does not vanish exponentially as the system size grows. The SKQD method constructs a set of \textit{Krylov states} by repeatedly applying the time-evolution operator to the reference state:
\begin{equation}
    \ket{\Psi_k} = e^{-iHkt} \ket{\Psi_{\text{init}}}, \qquad k = 0, 1, \ldots, K ,
\end{equation}
where $t$ is a small evolution time and $K$ determines the Krylov subspace dimension. Each $\ket{\Psi_k}$ can be interpreted as a snapshot of the quantum dynamics governed by $H$. Sampling measurement outcomes from these time-evolved circuits produces bitstrings corresponding to the basis states that dominate the true ground-state wavefunction.

The Hamiltonian is then projected into the linear span of the sampled bitstrings and classically diagonalized:
\begin{equation}
    H_{\text{eff}} = P^{\dagger} H P ,
\end{equation}
where $P$ is the matrix whose columns correspond to the sampled basis vectors. Solving the eigenvalue problem for $H_{\text{eff}}$ yields an estimate of the ground-state energy and a reconstruction of the approximate ground-state wavefunction in the sampled subspace. The accuracy of the method is controlled by the size of the subspace. The larger the subspace, the more accurate the ground state approximation becomes, at an increased quantum and classical runtime. A diagram of our workflow can be found in Fig.~\ref{fig:model_architecture}.

\subsection{Hybrid Quantum-Classical Training} \label{sec:Training}
The training of the model represented in the Pauli string basis in Eq.~(\ref{eq:bxo}) involves learning the coefficients $\{\beta^p,\ \chi_f^p, \ o_j^p \}$ by solving Eq.~(\ref{eq:train}).  As with any AI/ML training procedure, hybrid training requires choosing hyper-parameters $\mathcal{H}$, which include the optimizer and its parameters (batch size, learning rate and its schedule), SKQD parameters (unitary Krylov subspace dimensions, number of shots, evolution time step, a subspace selection strategy, and the number of energy terms in the gradient expansion in Eq.~(\ref{eq:gradients}).  An example of a training workflow, with explicit values of these parameters, is illustrated in Algorithm\ref{alg:one}. 

\RestyleAlgo{ruled}
\SetAlgorithmName{Algorithm 1}{Workflows}{List of Workflows}
\SetKwComment{Comment}{/* }{ */}
\SetKwInput{KwPauliSet}{Pauli String Set $P$}
\SetKwInput{kwHyperParams}{Hyper-parameters $\mathcal{H}$}
\SetKwInput{KwTrainingData}{Training Data $X$}
\SetKwInput{KwModelParameters}{Model Parameters $\mathcal{M}$}
\SetKwInOut{KwModelOutput}{Result (Output)}
\SetKwBlock{DoParallel}{do in parallel}{end}
\begin{algorithm}[hbt!]
\label{alg:one}
\caption{Example of a Supervised Training workflow}
\SetAlFnt{\small}
\KwTrainingData{400 data points with 10 features and 2 classes}
\KwPauliSet{Adjacent 2-local Pauli strings with X and Z used in this case}
\KwModelParameters{Coefficients $\{ \beta^p, \ \chi_f^p, \ o_k^p \}$ describing the bias operator, matrix configuration, and output operators}
\kwHyperParams{Batch Size = 25, Optimizer = Adam, Learning Rate = 0.1 reduced by 0.8x every 5 epochs, Unitary Krylov Dimension = 8, Number of Shots = 200, Evolution Time Step $\in [0.005,0.1]$, Number of Energy Terms $\in [2,60]$ }
\KwModelOutput{Trained model parameters $\mathcal{M}$, final bitstring pool}
\For{\text{epoch=1 to final\_epoch}}{
  \For{\text{batch=1 to final\_batch}}{
    \For{\text{item=1 to final\_batch\_element in parallel}} 
    {
    \begin{enumerate}
      \small{
      \item Create sparse Pauli operators representation using model's parameters;
      \item Based on resampling strategy, calculate bitstring pool using sampling strategy;
      \item With bitstring pool, calculate: $\{E_k\}_{k=0}^{\text{\# of energy terms}},\{\bra{\psi_0}P_i\ket{\psi_k}\}_{(k,i)=(0,1)}^{(\text{\# of energy terms,P)}}$;
      \item Calculate logits using \ref{eq:logits} and \ref{eq:bxo};
      \item Using \ref{eq:gradients} calculate gradients; and
      \item Using the optimizer strategy, update model parameters
      }
    \end{enumerate}
    }
  }
}
\end{algorithm}

Quantum processors are used only in Step 2, and require a sampling strategy that determines how a bistring pool is created, and a resampling strategy that determines when the pool is updated.  Examples of sampling strategies include a) "last pool" which uses the bitstring pool associated with the last datapoint whose Hamiltonian was processed, and b) "batch-size union", which uses the union of all the distinct bitstrings observed for a given batch.  Similarly, various resampling strategies are possible, including a) "beginning of epoch", which updates the bitstring pool at the beginning of each epoch, and b) "greedy epoch strategy" that does not resample as long as training loss declines.  In this study, the "batch-size union" strategy combined with the "beginning of epoch" resampling was observed to be sufficient to train the model up to 50 qubits. \\

\begin{figure}[!htbp]
\centering
\includegraphics[scale=0.8]{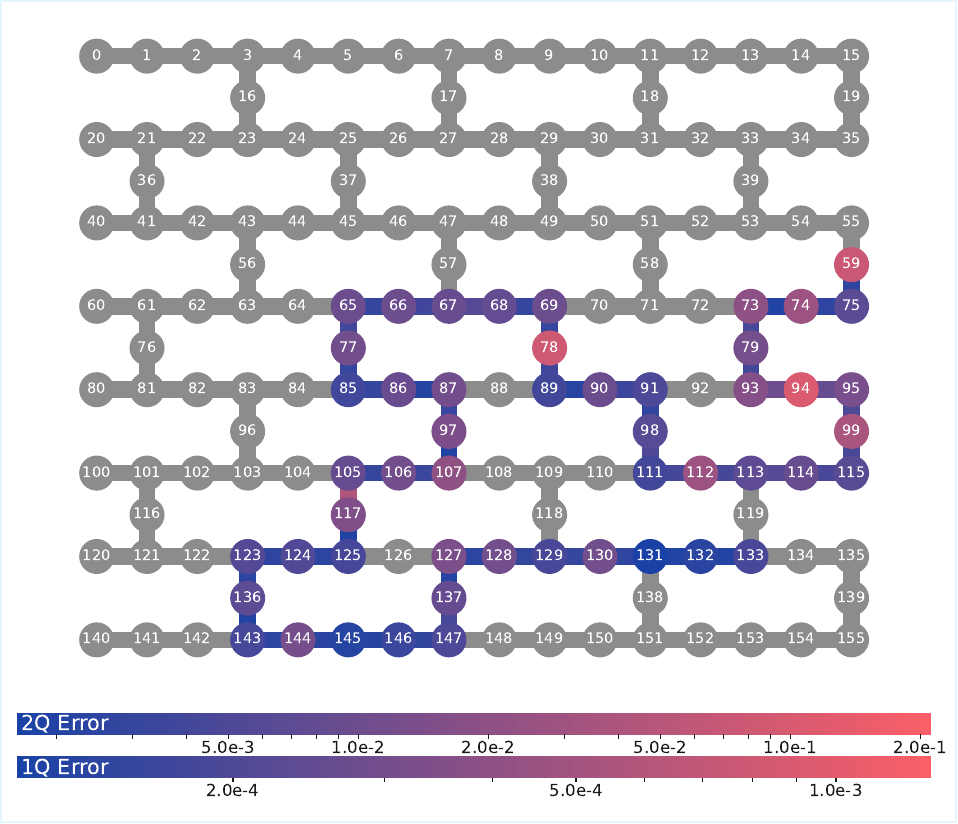}
\caption{\textbf{Device topology of $ibm\_fez$} There are 156 qubits in this processor and we selected 50 qubits out of these for our experiment. These qubits and their connected edges are highlighted and color coded based on single qubit and two qubit error rates learned using randomized benchmarking experiments.}
\label{fig:ibm_fez}
\end{figure}

\noindent \textit{Execution of quantum circuits on IBM processor}

As discussed earlier, a quantum computer is used to sample bitstrings as a subroutine of SKQD for computing the energy terms in the gradient calculation. We use $ibm\_fez$, a 156 qubit IBM heron processor depicted in Fig. \ref{fig:ibm_fez} with heavy-hex connectivity. In order to deal with noise that is present in our pre fault-tolerant devices, we employ a variety of techniques to improve the quality of bitstrings sampled:
\begin{itemize}
    \item Qubit selection is very important as evident from in-homogeneous error rates across the chip. We perform a layer fidelity experiment across the entire 156 qubit device. Layer fidelity is a benchmark that encapsulates the quality of circuits a particular processor is capable of running taking into account various metrics like qubit qualities, gate error rates, gate speed, etc. We down select a chain of 50 qubits using this layer fidelity estimate.
    \item We perform dynamical decoupling to suppress decoherence during idle periods within our circuit. 
\end{itemize}

\section{Results}
\label{sec:Results}

\begin{figure}[!htbp]
\centering
\includegraphics[scale=0.6]{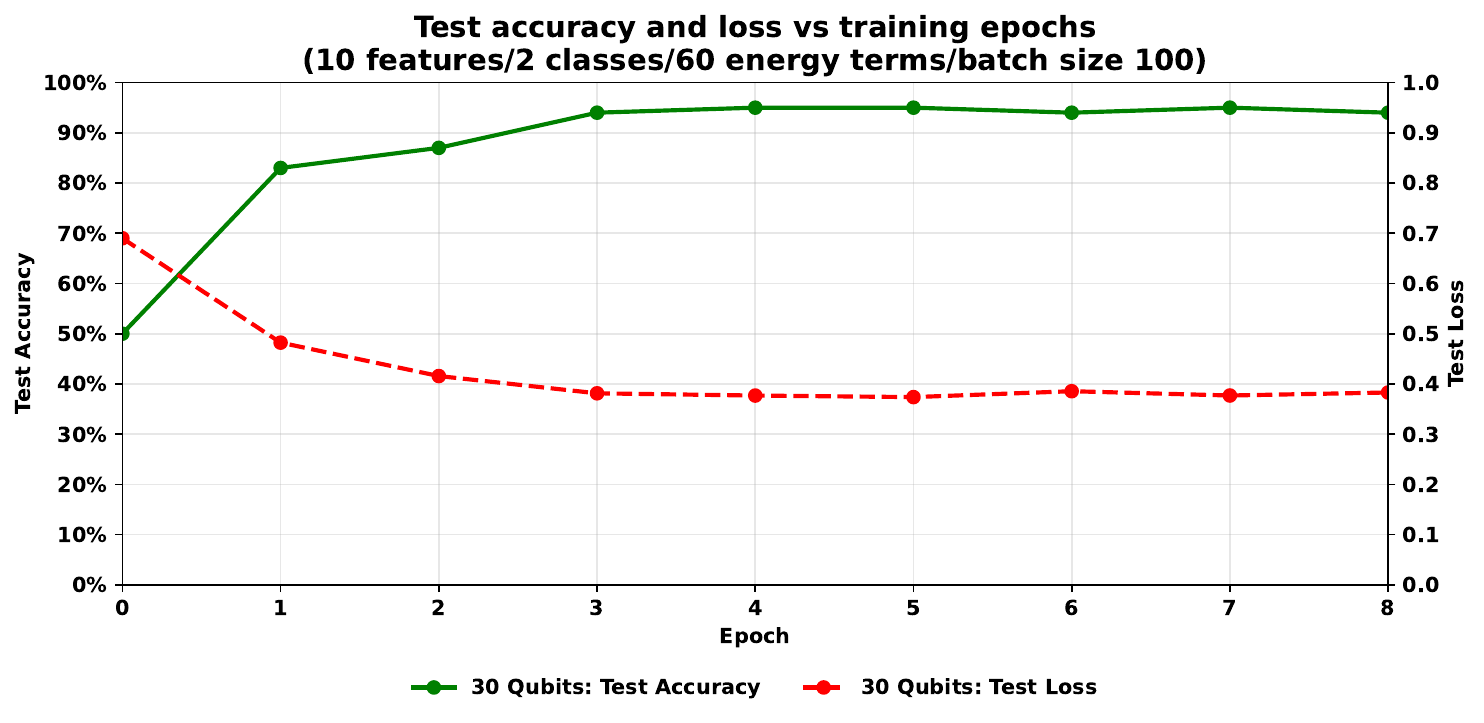} \\
\includegraphics[scale=0.6]{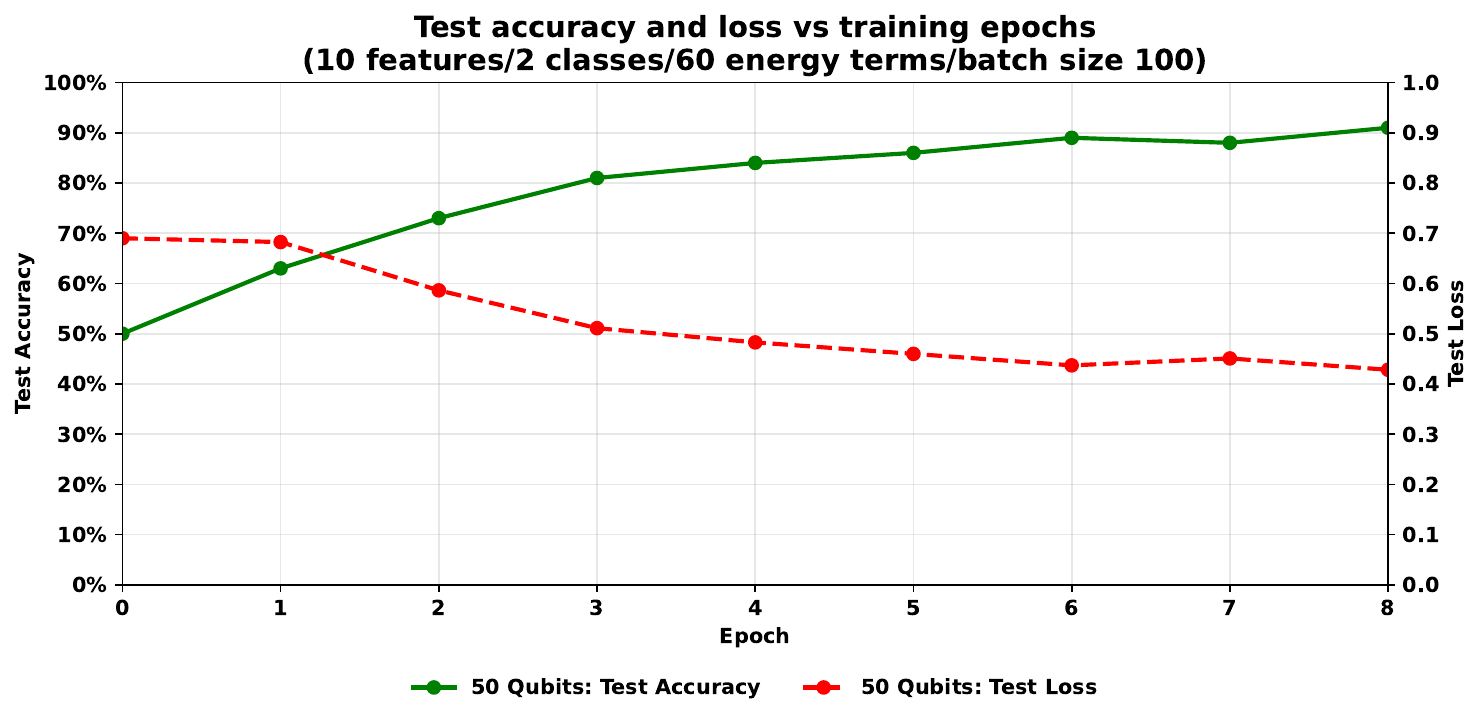}
\caption{Validation loss and accuracy vs training epochs on 30 and 50 qubits}
\label{fig:50qubit_accuracy_loss}
\end{figure}

In keeping with the spirit of data analysis and machine learning research, the experiments reported in this paper use a synthetic classification dataset generated using the \texttt{scikit-learn}/ SciPy ecosystem. Synthetic datasets provide full control over class structure, feature correlations, and noise levels, allowing a clear evaluation of algorithmic behavior without the confounding factors that are inherent in real-world data. The controlled setting enables assessment of the hybrid training procedure, and the efficacy of the SKQD-based ground-state estimation.  Our experiments consider a binary classification task with 500 data points, each with 10 features, using an 80/20 train-test split to evaluate training performance and generalization accuracy. It is used herein primarily as proof-of-concept to validate the proposed methodology and to place it in the context of conventional machine learning benchmarks.

Models were trained across system sizes ranging from 3 to 50 qubits under multiple hyperparameter configurations. Experiments at smaller sizes ($\le 18$) were performed on a quantum simulator and used for preliminary validation and for selection of an effective set of hyperparameters.  Larger sizes, $\ge 20$ qubits, utilized IBM's Heron quantum processors.  The results from the 30 and 50-qubit models are illustrated in Fig.~\ref{fig:50qubit_accuracy_loss}, where the test loss declines from approximately .69 to .42, and the accuracy climbs from 50\% to 91\%. This clearly demonstrates that the model trains successfully and is able to achieve high accuracy. The quantum circuits used to run the trotterization for SKQD on the 50 qubit model used over 1000 entangling two qubit operations (CZ). 

\begin{figure}[!htbp]
\centering
\includegraphics[scale=0.6]{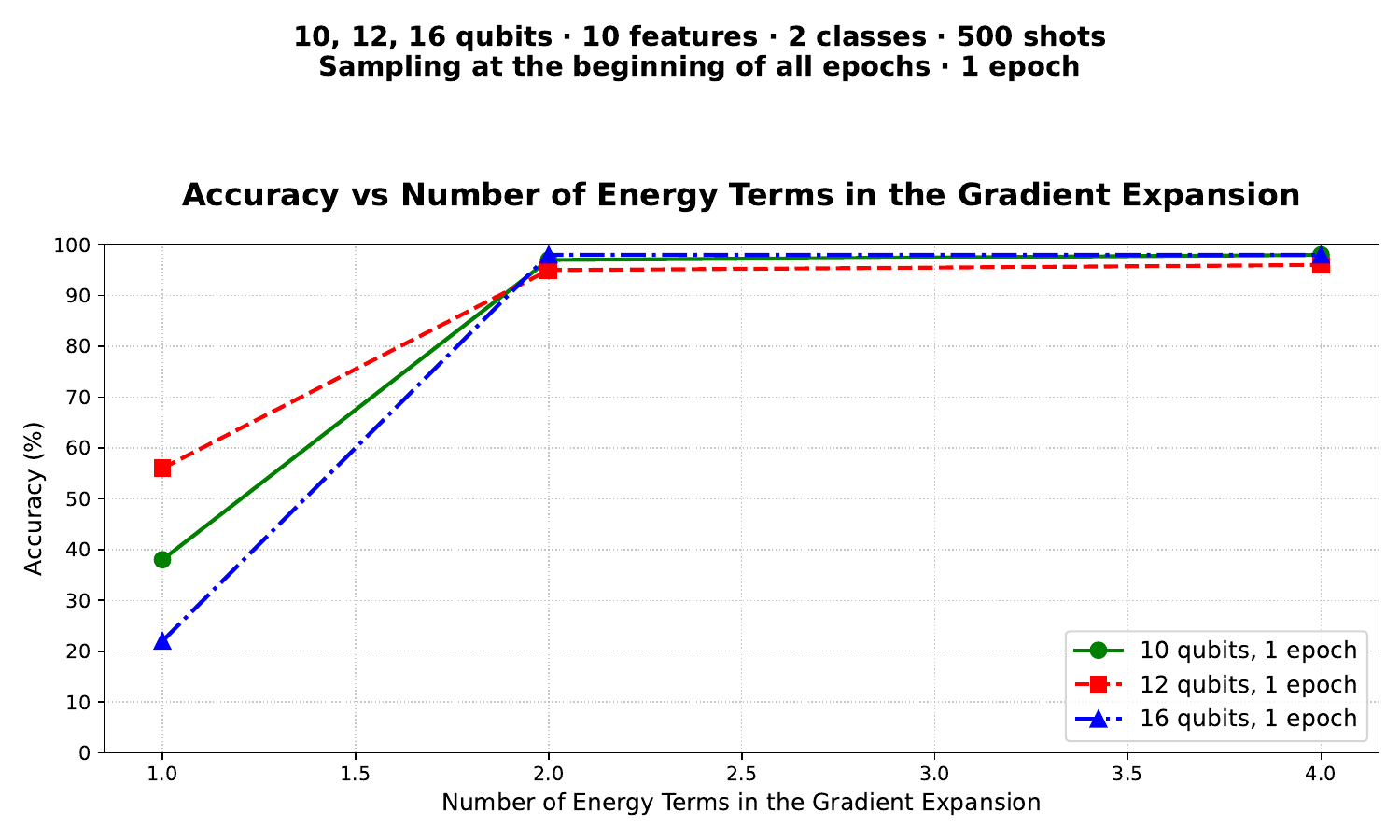} \\
\caption{Classical simulation of validation accuracy vs training epochs or number of energy terms }
\label{fig:classical_sim}
\end{figure}

Finally, we summarize our detailed numerical simulations aimed at understanding the behavior of this novel method, as shown in Fig. \ref{fig:classical_sim}. The accuracy of the gradient computed in Eq. (\ref{eq:gradients}) depends on the number of energy terms included. Since the classical computation time for this protocol grows with the number of terms, we benchmarked the validation accuracy of 10-, 12-, and 16-qubit models on a 10-feature, 2-class classification problem after one epoch of training, progressively adding more energy terms. Surprisingly, for this dataset, including just two energy terms is sufficient. While this observation may not generalize to all datasets or qubit configurations, it suggests that achieving high accuracy does not require an impractically large number of terms in the gradient calculation.



\section{Discussion}
\label{sec:DiscussionAndNextSteps}

In this work, we demonstrate successful training of a linear Hamiltonian-based machine learing model on current quantum processors with non-vanishing gradients observed on devices in the 30-50 qubits regime. Our approach utilizes sparse ground-state representations obtained from the SKQD algorithm, which enables learning within a low-dimensional subspace and may help mitigate optimization challenges such as barren plateaus \cite{larocca2025barren}. 

Hamiltonian-based encodings offer a compact and expressive alternative to existing data-embedding strategies, with the learned operators adapting to the structure of the ML task. In our experiments, we employ $2$-local operators, resulting in time-evolution circuits with shallow depth. Similar ideas could be applied to amplitude- or angle-based encodings, where classical pre-processing is typically used to reduce feature dimensionality before the data is embedded in a QML model. In contrast, our framework learns such an encoding directly as part of the algorithm. 

There are several important directions for future work. Further optimization of hyperparameters may improve training dynamics of these models and help identify more expressive operator ansatz. An important next step is to evaluate our approach on larger feature sets and more challenging datasets. A systematic comparison against classical methods would be crucial in identifying regimes where Hamiltonian-based encodings offer a genuine advantage.

\bibliographystyle{abbrv} 
\bibliography{ref}

@article{agliardi2024mitigating,
  title={Mitigating exponential concentration in covariant quantum kernels for subspace and real-world data},
  author={Agliardi, Gabriele and Cortiana, Giorgio and Dekusar, Anton and Ghosh, Kumar and Mohseni, Naeimeh and O'Meara, Corey and Valls, V{\'\i}ctor and Yogaraj, Kavitha and Zhuk, Sergiy},
  journal={arXiv:2412.07915},
  year={2024}
}

@article{liu2021rigorous,
  title={A rigorous and robust quantum speed-up in supervised machine learning},
  author={Liu, Yunchao and Arunachalam, Srinivasan and Temme, Kristan},
  journal={Nature Physics},
  volume={17},
  number={9},
  pages={1013--1017},
  year={2021},
  publisher={Nature Publishing Group UK London}
}

@article{deshpande2024dynamic,   title={Dynamic parameterized quantum circuits: expressive and barren-plateau free},   author={Deshpande, Abhinav and Hinsche, Marcel and Najafi, Khadijeh and Sharma, Kunal and Sweke, Ryan and Zoufal, Christa},   journal={arXiv:2411.05760},   year={2024} }

@article{tang2022dequantizing,
  title={Dequantizing algorithms to understand quantum advantage in machine learning},
  author={Tang, Ewin},
  journal={Nature Reviews Physics},
  volume={4},
  number={11},
  pages={692--693},
  year={2022},
  publisher={Nature Publishing Group UK London}
}

@article{aaronson2015read,
  title={Read the fine print},
  author={Aaronson, Scott},
  journal={Nature Physics},
  volume={11},
  number={4},
  pages={291--293},
  year={2015},
  publisher={Nature Publishing Group UK London}
}

@article{harrow2009quantum,
  title={Quantum algorithm for linear systems of equations},
  author={Harrow, Aram W and Hassidim, Avinatan and Lloyd, Seth},
  journal={Physical review letters},
  volume={103},
  number={15},
  pages={150502},
  year={2009},
  publisher={APS}
}

@article{larocca2025barren,
  title={Barren plateaus in variational quantum computing},
  author={Larocca, Martin and Thanasilp, Supanut and Wang, Samson and Sharma, Kunal and Biamonte, Jacob and Coles, Patrick J and Cincio, Lukasz and McClean, Jarrod R and Holmes, Zo{\"e} and Cerezo, Marco},
  journal={Nature Reviews Physics},
  pages={1--16},
  year={2025},
  publisher={Nature Publishing Group UK London}
}

@article{jerbi2024shadows,
  title={Shadows of quantum machine learning},
  author={Jerbi, Sofiene and Gyurik, Casper and Marshall, Simon C and Molteni, Riccardo and Dunjko, Vedran},
  journal={Nature Communications},
  volume={15},
  number={1},
  pages={5676},
  year={2024},
  publisher={Nature Publishing Group UK London}
}

@article{kandala2019error,
  title={Error mitigation extends the computational reach of a noisy quantum processor},
  author={Kandala, Abhinav and Temme, Kristan and C{\'o}rcoles, Antonio D and Mezzacapo, Antonio and Chow, Jerry M and Gambetta, Jay M},
  journal={Nature},
  volume={567},
  number={7749},
  pages={491--495},
  year={2019},
  publisher={Nature Publishing Group UK London}
}

@article{lanes2025framework,
  title={A Framework for Quantum Advantage},
  author={Lanes, Olivia and Beji, Mourad and Corcoles, Antonio D and Dalyac, Constantin and Gambetta, Jay M and Henriet, Loic and Javadi-Abhari, Ali and Kandala, Abhinav and Mezzacapo, Antonio and Porter, Christopher and others},
  journal={arXiv:2506.20658},
  year={2025}
}

@article{abanov2025QuantumGeometryData,
  title = {Quantum {{Geometry}} of {{Data}}},
  author = {Abanov, Alexander G. and Candelori, Luca and Steinacker, Harold C. and Wells, Martin T. and Busemeyer, Jerome R. and Hogan, Cameron J. and Kirakosyan, Vahagn and Marzari, Nicola and Pinnamaneni, Sunil and Villani, Dario and Xu, Mengjia and Musaelian, Kharen},
  journal = {arXiv:2507.21135},
  year = 2025,
  month = jul,
  urldate = {2025-11-29},
  archiveprefix = {arXiv}
}

@article{anschuetz2022QuantumVariationalAlgorithms,
  title = {Quantum Variational Algorithms Are Swamped with Traps},
  author = {Anschuetz, Eric R. and Kiani, Bobak T.},
  year = 2022,
  month = dec,
  journal = {Nature Communications},
  volume = {13},
  number = {1},
  pages = {7760},
  publisher = {Nature Publishing Group},
  issn = {2041-1723},
  doi = {10.1038/s41467-022-35364-5},
  urldate = {2025-11-29},
  copyright = {2022 The Author(s)},
  langid = {english}
}

@article{candelori2025RobustEstimationIntrinsic,
  title = {Robust Estimation of the Intrinsic Dimension of Data Sets with Quantum Cognition Machine Learning},
  author = {Candelori, Luca and Abanov, Alexander G. and Berger, Jeffrey and Hogan, Cameron J. and Kirakosyan, Vahagn and Musaelian, Kharen and Samson, Ryan and Smith, James E. T. and Villani, Dario and Wells, Martin T. and Xu, Mengjia},
  year = 2025,
  month = feb,
  journal = {Scientific Reports},
  volume = {15},
  number = {1},
  pages = {6933},
  publisher = {Nature Publishing Group},
  issn = {2045-2322},
  doi = {10.1038/s41598-025-91676-8},
  urldate = {2025-11-29},
  copyright = {2025 The Author(s)},
  langid = {english}
}

@article{caro2025QuantumCognitionMachine,
  title = {Quantum {{Cognition Machine Learning}} for {{Forecasting Chromosomal Instability}}},
  author = {Caro, Giuseppe Di and Kirakosyan, Vahagn and Abanov, Alexander G. and Busemeyer, Jerome R. and Candelori, Luca and Hartmann, Nadine and Lam, Ernest T. and Musaelian, Kharen and Samson, Ryan and Steinacker, Harold and Villani, Dario and Wells, Martin T. and Wenstrup, Richard J. and Xu, Mengjia},
  journal = {arXiv:2506.03199},
  year = 2025,
  month = jul,
  primaryclass = {q-bio},
  publisher = {arXiv},
  doi = {10.48550/arXiv.2506.03199},
  urldate = {2025-11-29},
  archiveprefix = {arXiv}
}

@article{havlicek2019SupervisedLearningQuantumenhanced,
  title = {Supervised Learning with Quantum-Enhanced Feature Spaces},
  author = {Havl{\'i}{\v c}ek, Vojt{\v e}ch and C{\'o}rcoles, Antonio D. and Temme, Kristan and Harrow, Aram W. and Kandala, Abhinav and Chow, Jerry M. and Gambetta, Jay M.},
  year = 2019,
  month = mar,
  journal = {Nature},
  volume = {567},
  number = {7747},
  pages = {209--212},
  publisher = {Nature Publishing Group},
  issn = {1476-4687},
  doi = {10.1038/s41586-019-0980-2},
  urldate = {2025-11-29},
  copyright = {2019 The Author(s), under exclusive licence to Springer Nature Limited},
  langid = {english}
}

@book{kato_perturbation_1995,
        address = {Berlin Heidelberg},
        title = {Perturbation {Theory} for {Linear} {Operators}},
        isbn = {978-3-540-58661-6},
        language = {English},
        publisher = {Springer},
        author = {Kato, Tosio},
        year = {1995},
}

@misc{kanno2023QSCI,
  title={Quantum-Selected Configuration Interaction: classical diagonalization of Hamiltonians in subspaces selected by quantum computers}, 
  author={Keita Kanno and Masaya Kohda and Ryosuke Imai and Sho Koh and Kosuke Mitarai and Wataru Mizukami and Yuya O. Nakagawa},
  year={2023},
  eprint={2302.11320},
  archivePrefix={arXiv},
  primaryClass={quant-ph},
  url={https://arxiv.org/abs/2302.11320}, 
}

@article{mayraz_recognizing_2002,
        title = {Recognizing handwritten digits using hierarchical products of experts},
        volume = {24},
        copyright = {https://ieeexplore.ieee.org/Xplorehelp/downloads/license-information/IEEE.html},
        issn = {01628828},
        url = {http://ieeexplore.ieee.org/document/982899/},
        doi = {10.1109/34.982899},
        language = {en},
        number = {2},
        urldate = {2025-12-07},
        journal = {IEEE Transactions on Pattern Analysis and Machine Intelligence},
        author = {Mayraz, G. and Hinton, G.E.},
        month = feb,
        year = {2002},
        pages = {189--197}
}

@misc{musaelian2024QuantumCognitionMachine,
  title = {Quantum {{Cognition Machine Learning AI Needs Quantum}}},
  author = {Musaelian, Kharen and Abanov, Alexander and Berger, Jeffrey and Candelori, Luca and Kirakosyan, Vahagn and Samson, Ryan and Smith, James and Villani, Dario},
  year = 2024,
  publisher = {Qognitive, Inc.},
  langid = {english}
}

@book{nielsen_quantum_2010,
        title = {Quantum computation and quantum information},
        url = {https://books.google.com/books?hl=en&lr=&id=-s4DEy7o-a0C&oi=fnd&pg=PR17&dq=info:--Y2fz5P81wJ:scholar.google.com&ots=NJ7JdmnwVu&sig=OCMzeTxcqa5hDQPBxSo0l6Ciu_E},
        urldate = {2025-12-07},
        publisher = {Cambridge university press},
        author = {Nielsen, Michael A. and Chuang, Isaac L.},
        year = {2010}
}

@article{Robledo2024SKQD, 
  author = {Javier Robledo-Moreno  and Mario Motta  and Holger Haas  and Ali Javadi-Abhari  and Petar Jurcevic  and William Kirby  and Simon Martiel  and Kunal Sharma  and Sandeep Sharma  and Tomonori Shirakawa  and Iskandar Sitdikov  and Rong-Yang Sun  and Kevin J. Sung  and Maika Takita  and Minh C. Tran  and Seiji Yunoki  and Antonio Mezzacapo },
  title = {Chemistry beyond the scale of exact diagonalization on a quantum-centric supercomputer},
  journal = {Science Advances},
  volume = {11},
  number = {25},
  pages = {eadu9991},
  year = {2025},
  doi = {10.1126/sciadv.adu9991},
  URL = {https://www.science.org/doi/abs/10.1126/sciadv.adu9991},
  eprint = {https://www.science.org/doi/pdf/10.1126/sciadv.adu9991},
}

@misc{rosaler2025SupervisedSimilarityHighyield,
  title = {Supervised Similarity for High-Yield Bonds - {{Risk}}.Net},
  author = {Rosaler, Joshua and Candelori, Luca and Kirakosyan, Vahagn and Musaelian, Kharen and Samson, Ryan and Wells, Martin T. and Mehta, Dhagash and Pasquali, Stefano},
  year = 2025,
  month = jun,
  urldate = {2025-11-29},
  howpublished = {https://www.risk.net/cutting-edge/7961606/supervised-similarity-for-high-yield-bonds},
  langid = {english}
}

@misc{samson2024QuantumCognitionMachine,
  title = {Quantum Cognition Machine Learning: Financial Forecasting - {{Risk}}.Net},
  shorttitle = {Quantum Cognition Machine Learning},
  author = {Samson, Ryan and Berger, Jeffrey and Candelori, Luca and Kirakosyan, Vahagn and Musaelian, Kharen and Villani, Dario},
  year = 2024,
  month = oct,
  urldate = {2025-11-29},
  howpublished = {https://www.risk.net/cutting-edge/7960053/quantum-cognition-machine-learning-financial-forecasting},
  langid = {english}
}

@misc{yu2025QuantumCentricAlgorithmSampleBased,
  title = {Quantum-{{Centric Algorithm}} for {{Sample-Based Krylov Diagonalization}}},
  author = {Yu, Jeffery and Moreno, Javier Robledo and Iosue, Joseph T. and Bertels, Luke and Claudino, Daniel and Fuller, Bryce and Groszkowski, Peter and Humble, Travis S. and Jurcevic, Petar and Kirby, William and Maier, Thomas A. and Motta, Mario and Pokharel, Bibek and Seif, Alireza and Shehata, Amir and Sung, Kevin J. and Tran, Minh C. and Tripathi, Vinay and Mezzacapo, Antonio and Sharma, Kunal},
  year = 2025,
  month = sep,
  number = {arXiv:2501.09702},
  eprint = {2501.09702},
  primaryclass = {quant-ph},
  publisher = {arXiv},
  doi = {10.48550/arXiv.2501.09702},
  urldate = {2025-11-29},
  archiveprefix = {arXiv}
}

\end{document}